\documentclass[letters,fleqn,usenatbib]{mnras}

\usepackage[T1]{fontenc}
\usepackage{ae,aecompl}

\usepackage{graphicx}	
\usepackage{amsmath}	
\usepackage{amssymb}	

\newcommand\degrees{^\circ}
\def\pasp{PASP} 
\def\aat{A\&AT}

\title[Peanut-shaped metallicity distributions in bulges]{Peanut-shaped metallicity distributions in bulges of edge-on galaxies: the case of NGC 4710.\thanks{Based on observations collected at the ESO La Silla-Paranal Observatory within MUSE science verification program 60.A-9307(A).}}

\author[Oscar A. Gonzalez]{
Oscar A. Gonzalez,$^{1}$\thanks{E-mail:oscar.gonzalez@stfc.ac.uk}
Victor P. Debattista,$^2$
Melissa Ness,$^3$
Peter Erwin,$^4$\newauthor
~Dimitri A. Gadotti$^5$
\\
  $^1$ UK Astronomy Technology Centre, Royal Observatory, Blackford Hill, Edinburgh, EH9 3HJ, UK\\
  $^2$ Jeremiah Horrocks Institute, University of Central Lancashire,
  Preston, PR1 2HE, UK \\
  $^3$ Max-Planck-Institut fur Astronomie, Konigstuhl 17, D- 69117 Heidelberg, Germany \\
  $^4$ Max-Planck-Insitut f\"{u}r extraterrestrische Physik, Giessenbachstrasse, 85748 Garching, Germany \\
  $^5$ European Southern Observatory, Ave. Alonso de Cordova 3107, Casilla 19, 19001, Santiago, Chile}

\date{Accepted XXX. Received YYY; in original form ZZZ}

\pubyear{2016}

\begin{document}
\label{firstpage}
\pagerange{\pageref{firstpage}--\pageref{lastpage}}
\maketitle

\begin{abstract}
Bulges of edge-on galaxies are often boxy/peanut-shaped (B/PS), and
unsharp masks reveal the presence of an X shape.  Simulations show
that these shapes can be produced by dynamical processes driven by a
bar which vertically thickens the centre.  In the Milky Way, which
contains such a B/PS bulge, the X-shaped structure is traced by the
metal-rich stars but not by the metal-poor ones.  Recently \citet{Debattista+16} interpreted this property as a result of the varying effect of the bar on stellar populations with different starting kinematics. This {\it kinematic fractionation}
model predicts that cooler populations at the time of bar formation go
on to trace the X shape, whereas hotter populations are more uniformly
distributed.  As this prediction is not specific to
the Milky Way, we test it with MUSE observations of the B/PS bulge in
the nearby galaxy NGC~4710.  We show that the metallicity map is more
peanut-shaped than the density distribution itself, in good agreement
with the prediction.  This result indicates that the X-shaped
structure in B/PS bulges is formed of relatively metal-rich stars that
have been vertically redistributed by the bar, whereas the metal-poor
stars have a more uniform, box-shaped distribution.
\end{abstract}




\section{Introduction}
\label{sec:intro}

The study of the bulges of spiral galaxies is of fundamental
importance for obtaining a complete understanding of the processes
involved in the formation and evolution of disc galaxies. Early galaxy
mergers are thought to be very efficient in producing spheroids in the
central regions of disc galaxies \citep{Brooks+16}. These components,
known as \textit{classical bulges}, would have similar properties to
scaled-down elliptical galaxies \citep{KorKen04, Hopkins+10}.
However, observationally, the central regions of
galaxies are commonly found to be dominated by stellar bars
\citep[e.g.][]{Lutticke+00}. Furthermore, disc galaxies seen edge-on
often show central bulges that appear boxy, or sometimes peanut- or
X-shaped. Numerical simulations provide clues to the nature of these
systems placing them in the family of secular products arising from
the vertical heating and thickening of bars. Observations have
confirmed the link of this kind of bulges to the presence of stellar
bars \citep[see][for a detailed review]{Athanassoula+16}. These
boxy/peanut-shaped (B/PS) bulges are produced either by the buckling
instability of bars \citep{raha+91} or by resonant trapping
\citep{Quillen+14}. B/PS bulges are very common
\citep{laurikainen+16}, and even our own Milky Way (MW) Galaxy has a
B/PS bulge \citep[][and references therein]{zoccali+16}.

Secular evolution via bar buckling readily produces bulges with the
kinematics \citep[e.g.][]{shen+10} and X-shaped morphology
\citep[e.g.][]{wegg-gerhard+13} of the MW's bulge.  However, more
detailed studies of the stellar populations of the MW's B/PS bulge has
produced unexpected results in terms of a vertical metallicity
gradient \citep[][]{zoccali+08, gonzalez+13, johnson+13}, an X-shape that is traced by the
more metal-rich stars, but not by the metal-poorer ones
\citep[e.g.][]{ness-abu+13, vasquez+13}, and a nearly axisymmetric distribution of
RR-Lyrae variables at $b\simeq 4\degrees$ \citep{dekany+13}.  These properties of the
MW's bulge seemed to require the additional presence of a classical bulge.
Recently \citet{Debattista+16} resolved this discrepancy in the
secular evolution scenario by demonstrating how bars are able to
produce a separation between different stellar populations during the
B/PS bulge formation.  Because this separation is induced by the
different initial in-plane kinematics (which are correlated with ages
and therefore with metallicity), they adopted the term
\textit{kinematic fractionation} to refer to it.  They showed that the
properties of the resulting B/PS bulges nicely follow the trends
observed in the MW.

Kinematic fractionation is not specific to the Milky Way.  One
testable prediction of kinematic fractionation is that B/PS bulges
observed in edge-on disc galaxies will show metallicity maps that
trace a peanut shape, analogous to the X-shape seen in the MW.  In
this context, NGC~4710 is one of the best laboratories to test this
prediction. NGC~4710, located at 16.9 \mbox{Mpc}, has a stellar mass of $\rm M_{\star} \sim 7 \times 10^{10} M_{\sun}$ \citep{Cappellari+13}, is classified as an exactly edge-on barred galaxy with a B/PS bulge \citep{buta+15}, and has no co-spatial large-scale classical bulge \citep{gadotti+12, gonzalez+16}. Furthermore, the arms of its X-shaped bulge structure are seen very prominently, as a consequence of a nearly side-on
orientation of the bar \citep{buta+15}. In this Letter we investigate
the stellar population properties of the B/PS bulge of NGC~4710,
directly testing the prediction of kinematic fractionation.

\section{Observations}

We obtained spectral and imaging coverage of NGC~4710 using the Multi
Unit Spectroscopic Explorer (MUSE) instrument installed on the Very
Large Telescope (VLT). The observations were taken as part of the MUSE
Science Verification observing run in June 2014 and their details can
be found in the study of the kinematics of NGC~4710 presented in Gonzalez et
al. (2016). MUSE \citep{bacon+10} is an optical wide-field
integral-field spectrograph that uses the image slicing technique to
cover a field of view (FOV) of $1\arcmin\times1\arcmin$ in wide-field
mode resulting in a spatial sampling of $ \rm
0.2\arcsec\times0.2\arcsec$ spaxels and resolving power from ${\mathrm
  R} \sim 2000$ at $4600$~\AA\ to ${\mathrm R} \sim 4000$ at $9300$~\AA\
. The full field is split up into 24 subfields that are fed into one
of the 24 integral field units (IFUs) of the instrument. Observations
of NGC~4710 were carried without adaptive optics at the nominal
wavelength range ($\rm 4650$~\AA\ to $9300$~\AA). The central
coordinates of the observed field ($\rm \alpha = 12h49m37.9s$, $ \rm
\delta=+15^{\circ}10'00.8''$, J2000) were optimised as described in
\citet{gonzalez+16} to cover the inner part of the B/PS bulge and to
extend the coverage to the expected limit of the bulge on the NW side
of the galaxy.

\section{Simulation}

In this study we investigate the stellar population properties of the
B/PS bulge of NGC~4710 by comparing them to a star-forming $N$-body
chemodynamical simulation which evolved to form a B/PS bulge. The
simulation was presented by \citet{cole+14} and \citet{ness+14}, while
a full description of the physical processes involved was given in
\citet{Debattista+16}. In the simulation a disc galaxy forms purely
out of gas cooling from a spherical corona that settles into a disc,
triggering continuous star formation. The galaxy forms a strong bar
after 2 Gyr and by 10 Gyr it has formed a B/PS bulge \citep{Debattista+16}.  Very important for our study is the fact that all stars in
the simulation form from gas, rather than being put by hand as typically
done in $N$-body simulations.  Thus the simulation was able to
follow the chemical evolution of all stellar populations.

\section{Stellar populations analysis}

\begin{figure}
\centering
{
\includegraphics[scale=0.65,angle=0,trim=0.0cm 0.5cm 0.0cm 1.5cm, clip=true]{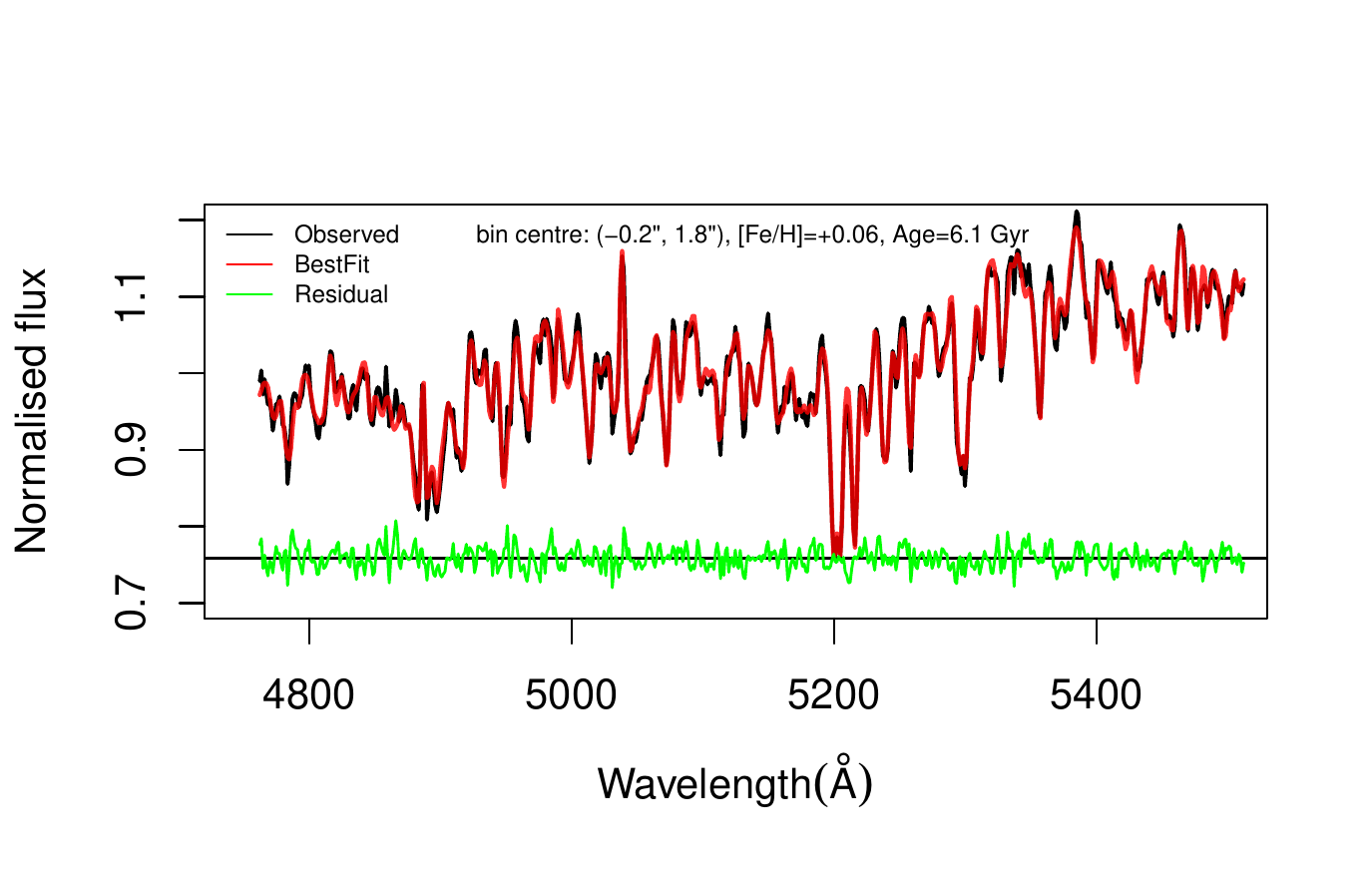}\\ 
\includegraphics[scale=0.65,angle=0,trim=0.0cm 0.5cm 0.0cm 1.5cm, clip=true]{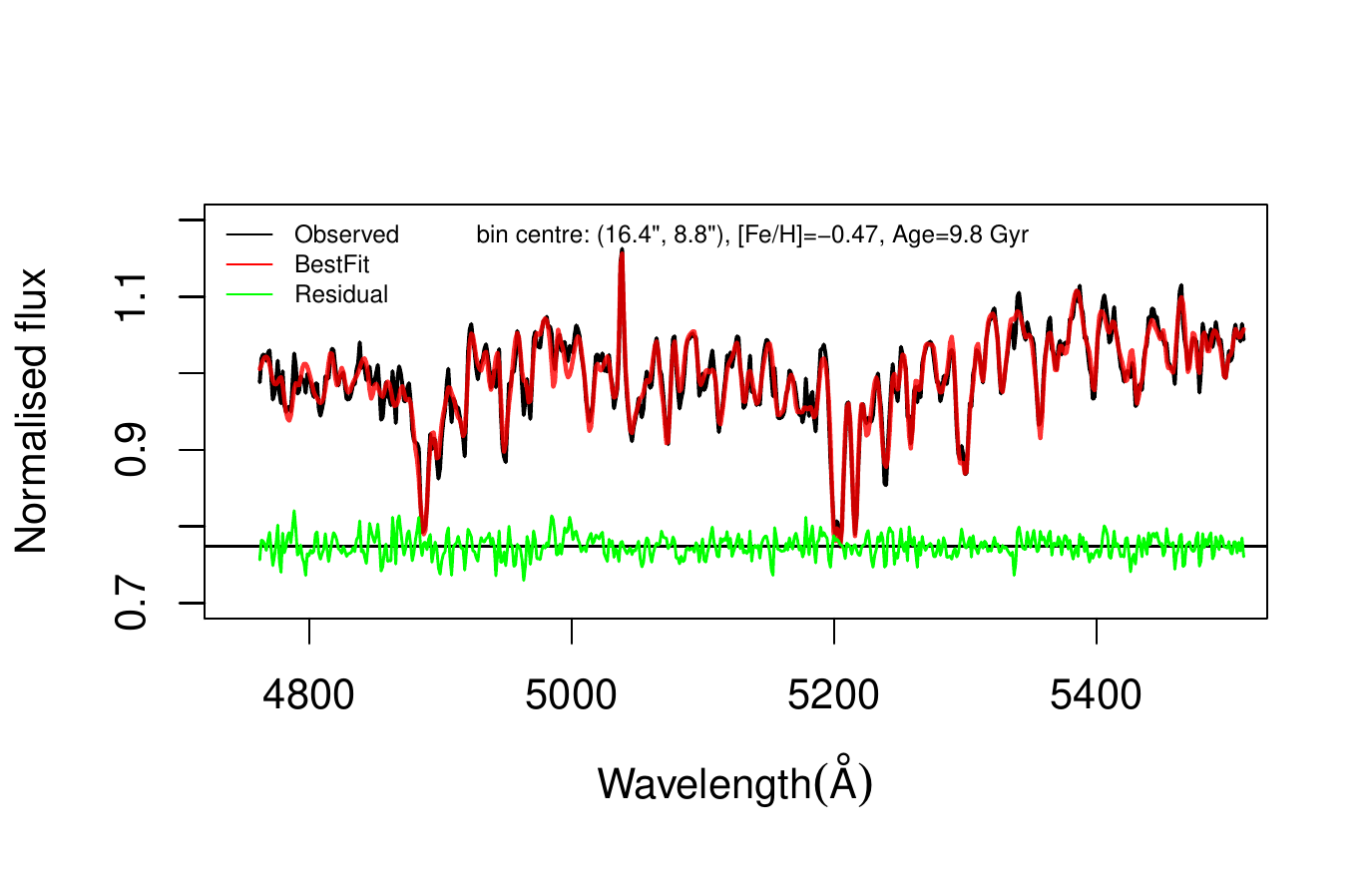}
}
\caption{Stellar population solutions of pPXF from the full spectral-fitting procedure for two representative regions of NGC~4710. The
  black lines show the observed, integrated spectra for two spatial
  bins with S/N$\sim$100 centred in the disc plane (upper panel) and in the
  outer bulge (lower panel) along the minor axis. The best-fit
  template spectra obtained from pPXF using the MIUSCAT libraries and
  emission line templates are shown as red lines. The 
  residuals from the fits for each case are shown in green. }
\label{4710_fit}
\end{figure}

\begin{figure*}
\centering
\resizebox{\hsize}{!}{
\includegraphics[scale=0.35,trim=0.0cm 0.0cm 2.5cm 0cm, clip=true]{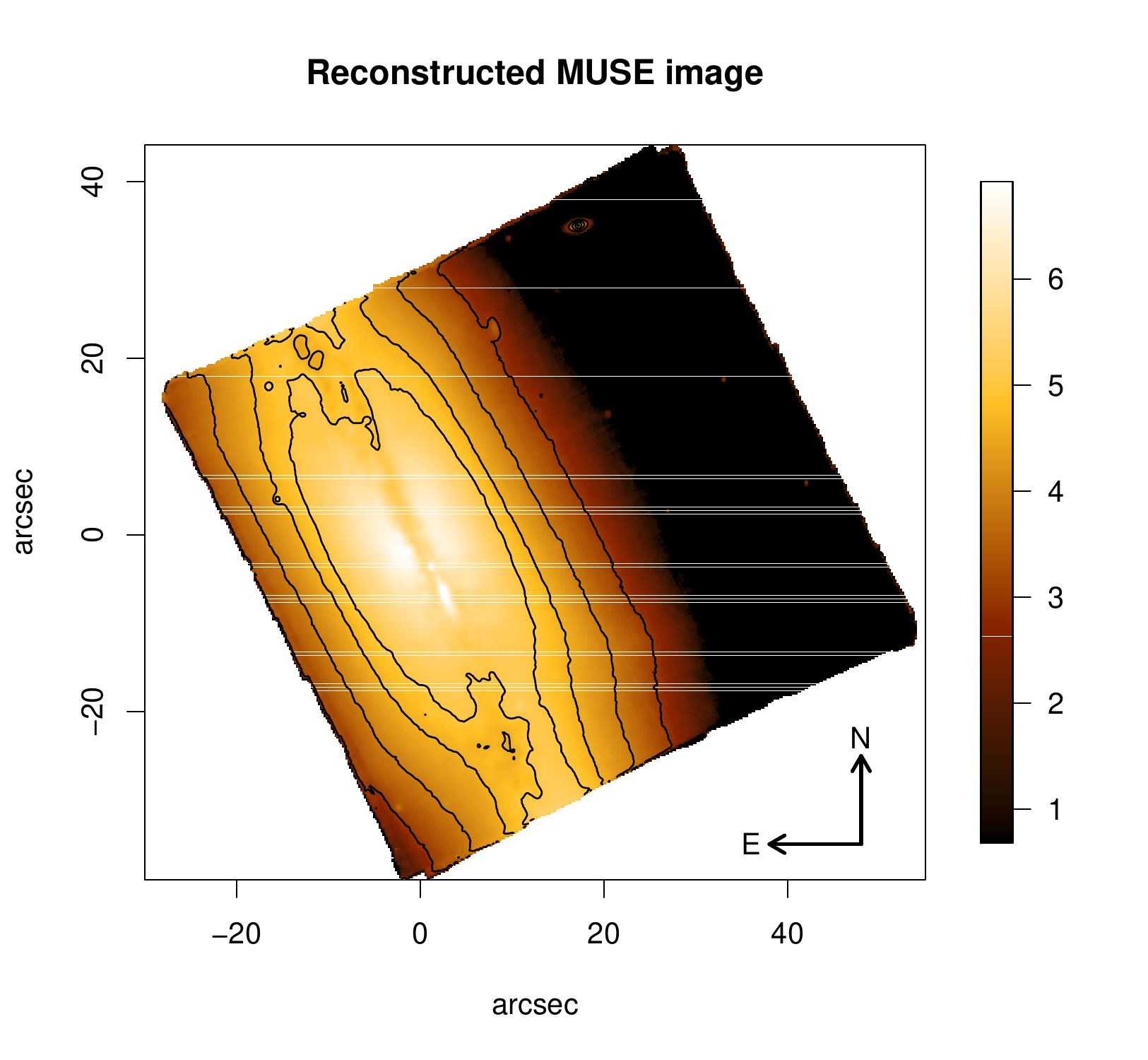} 
\includegraphics[scale=0.35]{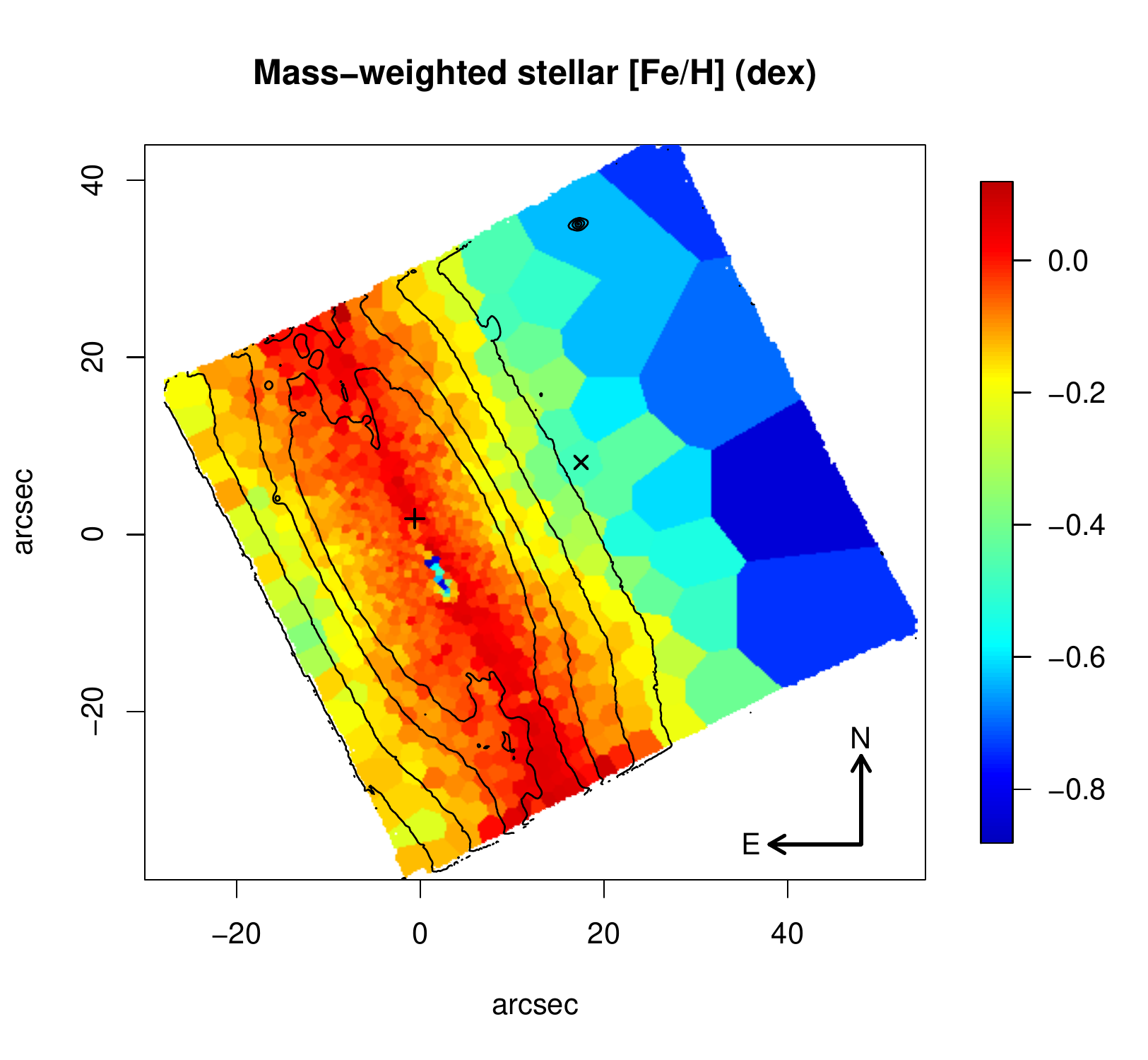}
\includegraphics[scale=0.35]{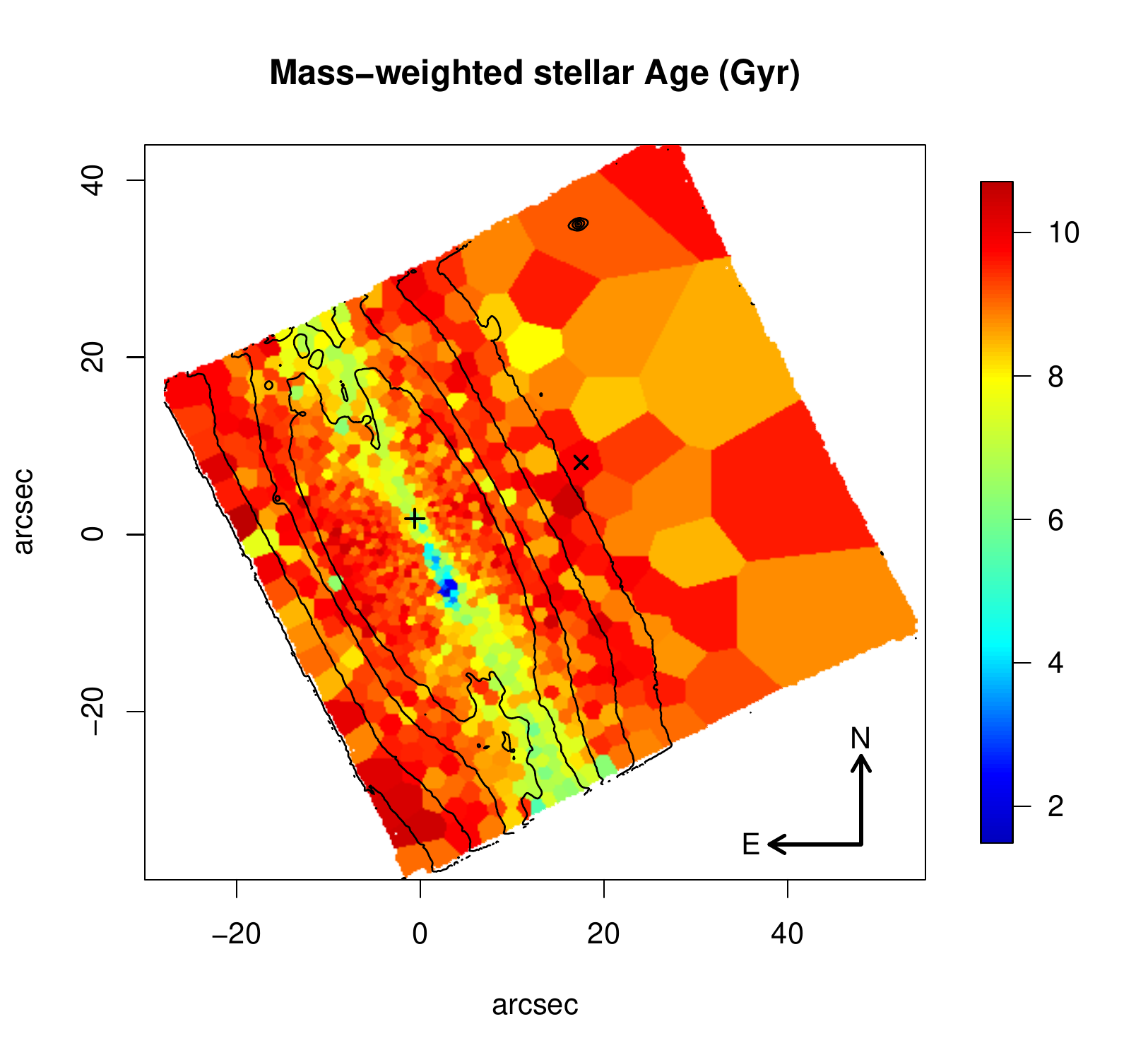}
}
\caption{The left panel shows the reconstructed flux map and isophote
  contours of the MUSE observations of NGC~4710 in arbitrary flux
  units. The mass-weighted stellar population maps of NGC~4710
  resulting from the regularised spectral fitting using pPXF are shown
  in the middle and right panels for metallicity and age,
  respectively. Isophote contours are over-plotted for reference to
  the regions mapped in the reconstructed image. The locations of the spectra displayed in the upper and lower panels of Fig.~\ref{4710_fit} are shown as $+$ and $\times$ symbols, respectively}.
\label{4710_map}
\end{figure*}

\citet{gonzalez+16} used the MUSE observations of NGC~4710 to map the
kinematics of its bulge. They used the Penalised Pixel Fitting routine
pPXF\footnote{http://www-astro.physics.ox.ac.uk/$\sim$mxc/software/}
\citep{capellari-emsellem+04} to evaluate the galaxy's stellar
kinematics by fitting the Single Stellar Population (SSP) MILES
library of \citet{vazdekis+10} to the observed spectra in pixel space
using a maximum penalised likelihood method. The line-of-sight (LOS) mean velocity and velocity dispersion were obtained by fitting templates to the observed spectra in each spatial bin.

In this work we extend the analysis of \citet{gonzalez+16} to
investigate the stellar populations of the bulge of NGC~4710. We
perform full-spectrum fitting with pPXF using the MIUSCAT
library\footnote{http://miles.iac.es/pages/stellar-libraries/miles-library.php}
of SSP models \citep{Vazdekis+12}. The SSP models from the MIUSCAT
library have been computed theoretically from the MILES and CaT
empirical spectra \citep{Cenarro+01} and were preferred for our study
as they have been shown to accurately reproduce the continuum and line
profiles, which are critical for the analysis of stellar
populations. We select a grid with ages ranging from 1 to 14.1 Gyr and
metallicities\footnote{We use the MIUSCAT \textit{base models}, which are computed using stars selected on the basis of their [Fe/H] metallicity. Therefore, these models follow the abundance pattern of the Milky Way.} ranging from -1.31 to +0.22 dex, computed using a \citet{salpeter+55} IMF (unimodal with slope 1.3) and based on the
Padova isochrones from \citet{Girardi+00}.

We follow the same methodology as in \citet{gonzalez+16} and use pPXF
to find the best fitting template spectrum over each spatial bin over
the wavelength range ($\rm 4750$~\AA\ to $\rm 5550$~\AA) to avoid
regions potentially contaminated by residuals of strong sky emission
lines. This spectral region has been used in similar analysis by the
SAURON \citep{bacon+01} and ATLAS$\rm^{3D}$ \citep{Cappellari+11}
surveys, as it includes strong stellar population indicators such as
H$\rm \beta$ (sensitive to age) as well as Mgb and Fe5250 (sensitive
to metallicity).

Gas emission lines are not masked, unlike in the earlier kinematic
analysis \citep[][]{gonzalez+16}, in order to include the absorption
H$\rm \beta$ line in the analysis. Instead, we perform a two component
fit in pPXF including the MIUSCAT templates as well as a set of
Gaussian emission templates. Further, we allow pPXF to treat the fit
of the emission lines template as a different kinematic component to
account for possible decoupled kinematics between gas and stellar
components. In order to accurately perform the two-component fit we
increase the binning with respect to the kinematic study (minimum
S/N=50) by imposing a minimum S/N of 100 in each spatial bin using the
Voronoi-binning method \citep{capellari-copin+03}.

We use only multiplicative polynomials to account for flux calibration
errors without including any additive polynomials. As described in
\citet{capellari-emsellem+04}, using additive polynomials is
recommended when fitting the LOS velocity distribution but they should
be avoided when inferring stellar populations as they can affect the
spectral line strength and bias the metallicity and age calculations.

We set pPXF to find an optimal spectral fitting solution based on the
weights applied to each template spectrum from the MIUSCAT library in
the age-metallicity grid using a regularisation parameter. As
described in \citet{Guerou+16}, the statistical
consistency of the best-fit solution to the data can be optimised by
increasing the regularisation parameter until the $\chi^2$ value of
the template fit exceeds that of the un-regularised fit by
$\sqrt{2N}$, where N is the number of spaxels. In this way we ensure
that the regularisation procedure constrains the solution such that
the weights assigned to neighbouring age and metallicity templates
changes smoothly while being fully consistent with our observed
spectrum.

\begin{figure*}
\centering
\includegraphics[scale=0.35, trim=0.0cm 0.0cm 2.5cm 0cm, clip=true]{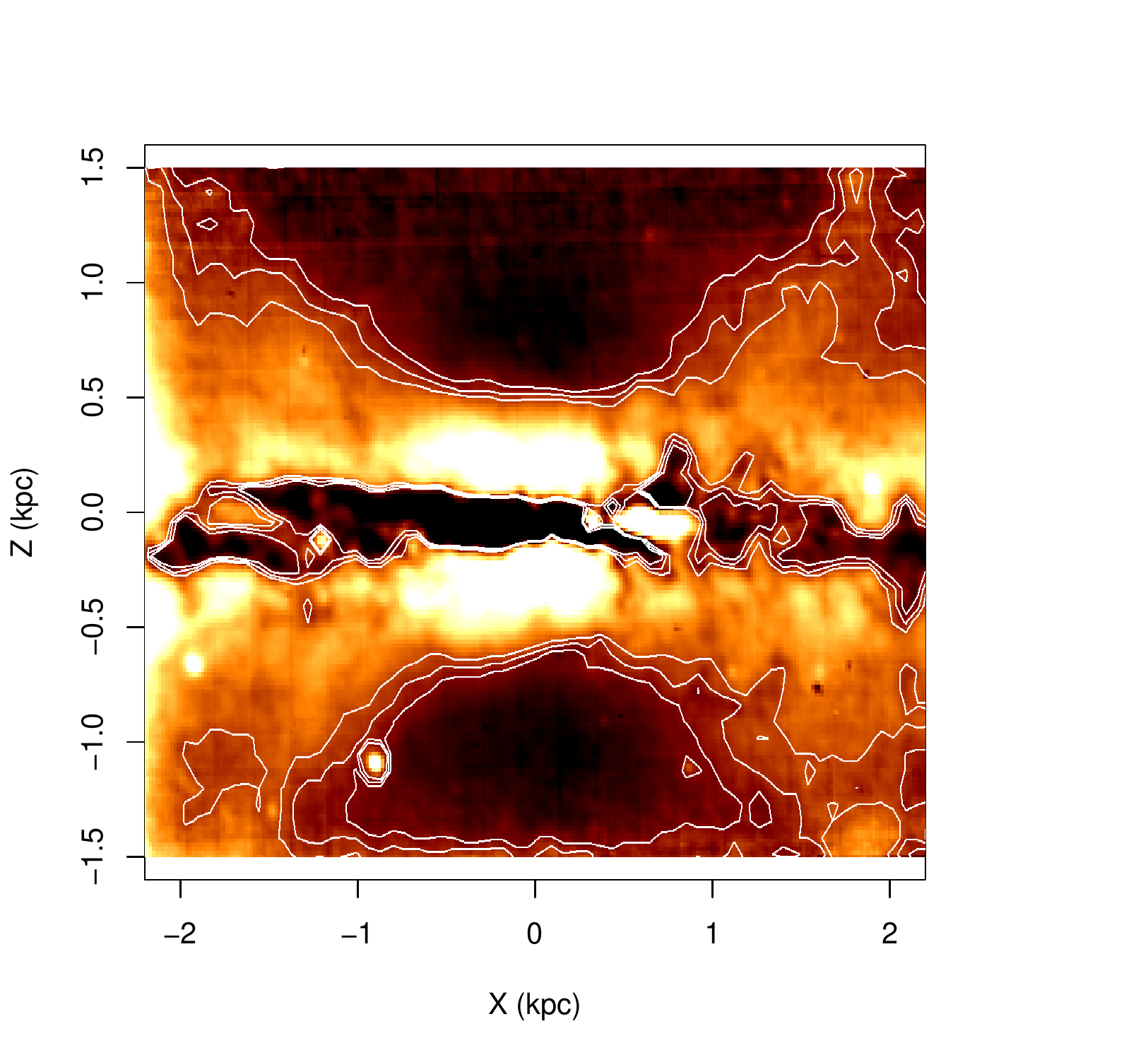}
\includegraphics[scale=0.35]{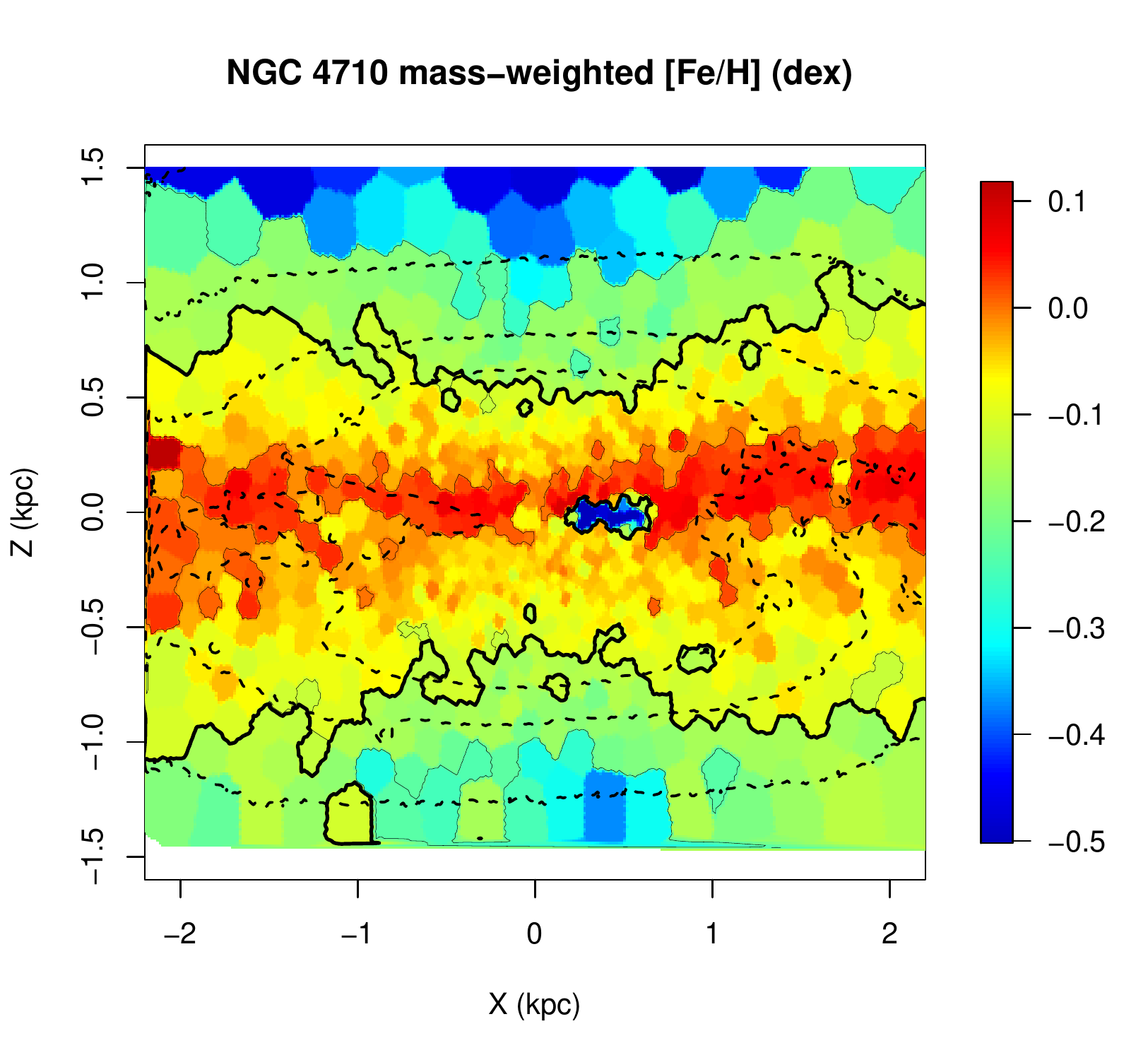} 
\includegraphics[scale=0.35]{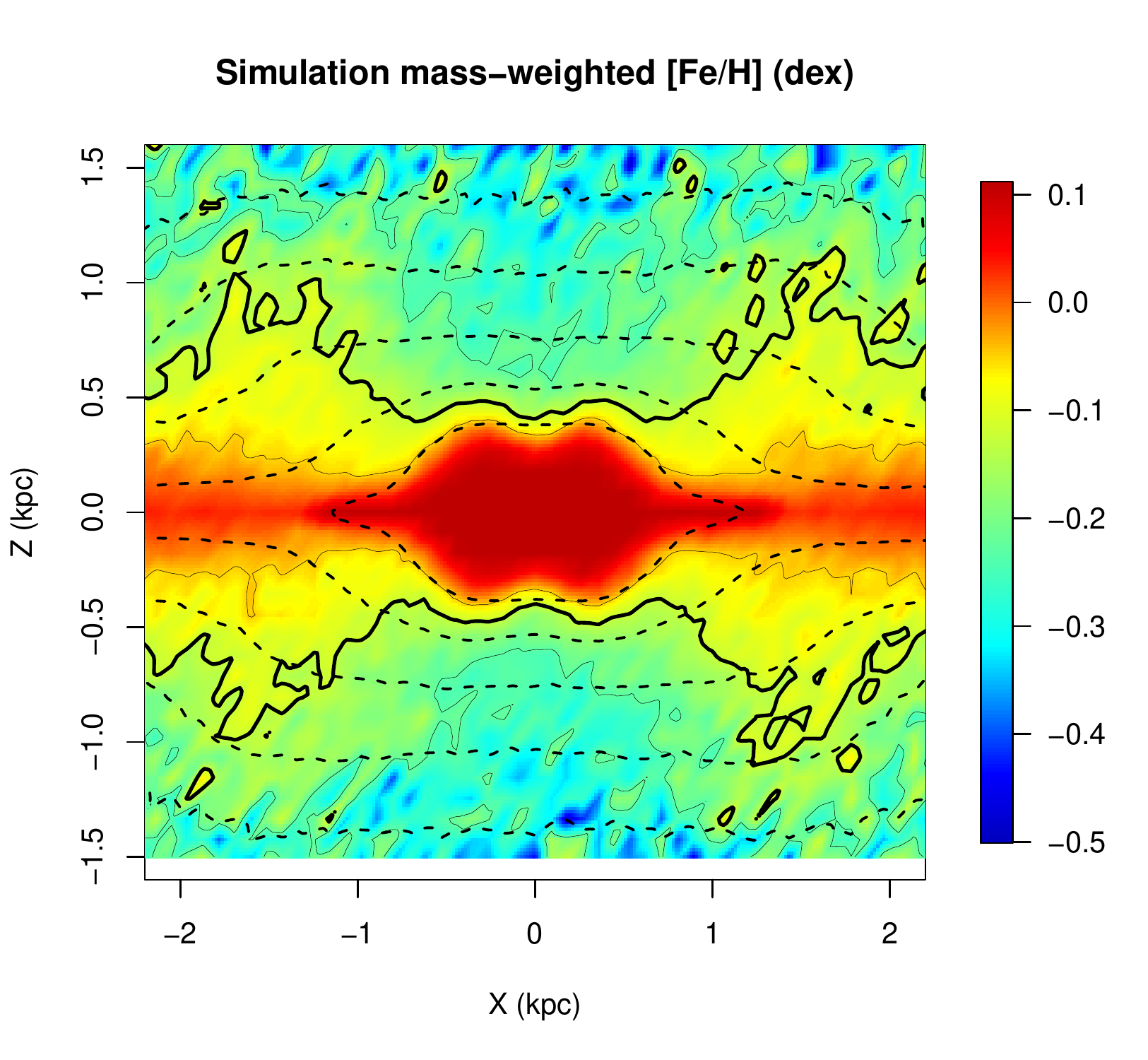}
\caption{The left panel shows the unsharp-masked MUSE image of
  NGC~4710 obtained by dividing the original image by a smoothed one using a Gaussian filter with $\rm \sigma=15$ pixels. The middle panel shows the mass-weighted metallicity map of NGC~4710. Dashed contour lines are stellar surface-brightness isophotes, as in Fig.~\ref{4710_map}. The right panel shows the metallicity map of the simulation from \citet{Debattista+16}. The density contours of the simulation are also shown in the right panel as dashed lines. The spatial coordinates of the simulation were scaled by a factor 1.2 and the $\rm [Fe/H]$ values in the simulation were decreased by 0.35 dex for comparison. The colour scale saturates at $\rm [Fe/H] = 0.1$ dex for visualisation. We assume a distance to NGC~4710 of 16.9 Mpc. The peanut shape of the metallicity maps is outlined by the boldfaced contour at $\rm [Fe/H]=-0.1$ for both NGC~4710 and the simulation.}
\label{4710_models}
\end{figure*}


The final representative stellar population of the observed integrated
spectrum for each bin is then obtained by calculating a weighted
average of the metallicities and ages of the grid using the weights
calculated by pPXF. Figure~\ref{4710_fit} shows the best fitting
solution for two spatial bins of the MUSE cube. The MIUSCAT template
models are normalised to a star of one solar mass so the resulting
mean age and metallicity values are mass-weighted. We estimate the
uncertainties in our measurements by assuming symmetry with
respect to the minor axis of the galaxy and evaluating the standard deviation of the [Fe/H] difference at symmetric locations. We find a typical error of $\sim0.03$ dex in
the mean [Fe/H] and 0.8 Gyr for the mean stellar age. These values are
expected to be slightly larger in the outermost bins that are located
furthest from the mid-plane.

\subsection{Age and metallicity maps of NGC~4710}

Figure~\ref{4710_map} shows the resulting mass-weighted 
metallicity and age maps of the bulge of NGC~4710. The maps suggest the
presence of a thin disc component of near-solar
metallicity with a mean age of 5-6 Gyr and they show that there is a
clear vertical metallicity gradient. We note that there is
a region near the mid-plane (but clearly not in the centre of the
galaxy) dominated by younger stellar populations (2 Gyr) than the rest
of the disc, which coincides with a flux excess seen in the
reconstructed image. In \citet{gonzalez+16} this region was identified
as a stellar population located in the foreground disc because of its
distinct kinematics and strong gas emission lines. The fact that the
stellar population in this region is also very young and metal-poor
with respect to the rest of the galaxy supports this conclusion and,
furthermore, suggests that this is a recently triggered star forming
region.

At larger heights from the plane, in the region dominated by the B/PS
bulge, the mean age of the stellar population quickly becomes older,
of the order of 8-9 Gyr, while the mean metallicity decreases to
$\rm [Fe/H]=-0.2$. Beyond this point, the age map is consistent with a
predominantly old population with a mean age of 10 Gyr while the
vertical metallicity gradient extends to $\sim35\arcsec$ ($\sim$2.0
kpc at 16.9 \mbox{Mpc}) from the plane of the galaxy, reaching a
metallicity of $\rm [Fe/H]=-0.8$ dex in the region of the outer
bulge and halo.

The most striking feature of the mass-weighted metallicity map of
NGC~4710 is its peanut shape, that is, the vertical metallicity
gradient has a different pattern along the minor axis than at the
sides of the B/PS bulge. The metal-rich population extends to larger
heights away from the plane at intermediate distances ($\sim1.5$~kpc) from the center
than it does along the minor axis. We refer to this gradient pattern
as the peanut-shaped metallicity map of NGC~4710.

\subsection{The origin of the population gradients in NGC~4710}

Simulations show that B/PS bulges can be produced by secular dynamical
processes driven by a bar. Bulges formed in this way often show an
underlying X-shaped structure in unsharp-mask images. This is the case
for NGC~4710, as shown in Figure~\ref{4710_models}, where the arms of
the X can be clearly identified. The unsharp-masked MUSE image was
constructed dividing the original image by a smoothed one that is obtained from convolving the original image with a Gaussian with $\rm \sigma=15$ pixels ($\rm 3\arcsec$). Figure~\ref{4710_models} also shows a more detailed view of
the mass-weighted metallicity map which appears more
pinched/peanut-shaped than the flux distribution off the mid-plane.

\citet{Debattista+16} predicted that kinematic fractionation
produces a B/PS bulge which has a metallicity distribution that is
even more peanut-shaped than the density itself.  The metallicity map
of their simulation is shown in the right panel of
Figure~\ref{4710_models}\footnote{For ease of comparison, all
  metallicities in the simulation are shifted by $-0.35$ dex.  Because
  here we are interested in the trends, not the absolute values, this
  shift has no effect on our analysis.}.  This shows a
remarkable qualitative similarity to the metallicity map of NGC~4710,
both in terms of shape and of vertical metallicity gradient.  This
metallicity distribution in the bulge is a consequence of the
anti-correlation between in-plane velocity dispersion and
metallicity at the time of bar formation. The metal-rich populations with low radial velocity dispersion are driven by
the bar to a vertically thick peanut-shaped bulge, whereas the hotter,
more metal-poor populations become vertically thicker with a more
boxy, rather than peanut, shape.

There is also a separate inner peanut-shaped, higher-metallicity region at $\rm r \la 0.5$ kpc
in the simulation ($\rm [Fe/H]>0.1$ dex region in the right panel of Fig.~3) which
is not seen in NGC~4710. However, as this is the region dominated by the (inner)
disc, this feature does not affect our conclusions.

The detection of the strong peanut-shape in the metallicity
distribution of the B/PS bulge of NGC~4710, which is also homogenously
old, represents strong evidence in favour of kinematic fractionation
having shaped its bulge.

\section{Conclusions}

We have shown for the first time, using VLT-MUSE IFU observations of
the edge-on galaxy NGC~4710, that at least some B/PS bulges show a
distinct peanut shape in the \textit{metallicity} distribution, with
the peanut shape more pronounced than in the density itself.  This
behaviour was predicted by the kinematic fractionation model of B/PS
bulge growth \citep{Debattista+16} and is a direct consequence of
the redistribution of stellar orbits by the bar based on their
in-plane velocities at the time of bar formation.

The peanut-shaped bulge metallicity distribution in NGC~4710 is homologous with
the X-shaped structure of the Milky Way's bulge, which is traced by
the double-peaked line-of-sight distribution of metal-rich stars, but
not by metal-poor ones \citep{ness+12, vasquez+13}.  The fact that
kinematic fractionation is able to produce vertical metallicity
gradients as strong as those observed in NGC~4710, and comparable to
those in the Milky Way \citep{gonzalez+16}, implies that vertical
metallicity gradients in B/PS bulges do not require an additional,
accreted bulge component (i.e. a classical bulge).  The absence of a
classical bulge in NGC~4710, as determined from photometric fitting,
supports the interpretation that the Milky Way also lacks a classical
bulge.

Therefore, the properties of the Milky Way's bulge, once considered
puzzling, now appear a rather natural outcome of purely secular
evolution.  We have shown here that such evolution also occurred in
NGC~4710, hinting that many of the Milky Way bulge's properties may be
very common amongst B/PS bulges.  Our results suggest a number of
fruitful next tests of kinematic fractionation.  Most obviously, the
number of bulges with studied metallicity maps needs to be expanded,
including in non-B/PS bulges. Kinematic fractionation also
predicts that B/PS bulges should exhibit a peanut-shape and vertical
gradient also in the age \citep{Debattista+16}. The age resolution of our spectra is not sufficient at the old ages we find in NGC~4710 to detect these variations in the age. Searching for these features in a larger sample of B/PS bulges would provide additional support that kinematic fractionation has sculpted the B/PS bulges observed today.

\section*{Acknowledgements}

We are grateful to Adrien Guerou for great advice and comments for the analysis of MUSE cubes and Ruben Sanchez-Janssen for useful discussions. VPD is supported by STFC Consolidated grant \#ST/M000877/1.  MN is funded by the European Research Council under the European Union's
Seventh Framework Programme (FP 7) ERC Grant Agreement \#321035.  We
acknowledge support from the ESF Exchange Grant (number 4650) within
the framework of the ESF Activity entitled 'Gaia Research for European
Astronomy Training'.  The star-forming simulation used in this paper
was run at the High Performance Computing Facility of the University
of Central Lancashire. VPD wishes to express his sincere thanks to the Pauli
Institute, the University of Z\"urich, and MPIA Heidelberg for hosting his sabbatical during which important work for this paper was completed. 




\bibliographystyle{mnras}
\bibliography{mybiblio_4710pop}








\bsp	
\label{lastpage}
\end{document}